\documentclass[double]{article}
\usepackage{times}
\usepackage{algorithm}
\usepackage{algorithmic}
\usepackage{wrapfig}
\usepackage{verbatim}
\usepackage{amsmath}
\usepackage{graphicx}
\usepackage{subcaption}
\usepackage[active]{srcltx}
\usepackage{color}
\usepackage{lineno}
\usepackage{dirtytalk}
\usepackage{amsthm}
\usepackage{authblk}
\usepackage{listings}

\usepackage{epsfig,graphicx,amsfonts}
\usepackage{amsmath,amsthm,amssymb}
\usepackage{xcolor}

\usepackage{adjustbox}
\usepackage{multirow}
\usepackage{setspace}
\usepackage{hyperref}
\usepackage{comment}

\long\def\comment #1\commentend{}

\begin{document}

\title{\Large A Computational Model For Individual Scholars' Writing Style Dynamics}
\author{Teddy Lazebnik$^{1*}$, Ariel Rosenfeld$^{2}$\\
\(^1\) Department of Cancer Biology, Cancer Institute, University College London, London, UK\\
\(^2\)  Department of Information Science, Bar Ilan University, Ramat-Gan, Israel\\
\(*\) Corresponding author: t.lazebnik@ucl.ac.uk
}

\maketitle 

\date{ }

\begin{abstract}
A manuscript's writing style is central in determining its readership, influence, and impact. Past research has shown that, in many cases, scholars present a unique writing style that is manifested in their manuscripts. In this work, we report a comprehensive investigation into how scholars' writing styles evolve throughout their careers focusing on their academic relations with their advisors and peers. Our results show that scholars' writing styles tend to stabilize early on in their careers -- roughly their 13th publication. Around the same time, scholars' departures from their advisors' writing styles seem to converge as well. Last, collaborations involving fewer scholars, scholars from the same gender, or from the same field of study seem to bring about greater change in their co-authors' writing styles with younger scholars being especially influenceable. \\
\noindent
\textbf{Keywords:} Academic writing style, Scientific communication, Computational linguistics, time-dependent mathematical graphs.
\end{abstract}


\pagestyle{myheadings} \markboth{Draft:  \today}{Draft:  \today}
\setcounter{page}{1}

\section{Introduction}
\label{sec:background}
Academic publications play a crucial role in the advancement and dissemination of knowledge and information \cite{intro_1}. 
In addition to many discipline-specific factors, such as scientific originality and validity, the way in which the manuscript is written, also known as the manuscript's \textit{style}, is pivotal in determining its readership, influence, and impact \cite{intro_2,intro_3,intro_4,journal_2,journal_4}. 
Specifically, a manuscript that is well-written, clear, easy to understand, and follows a logical flow and structure usually results in shorter reviewing time, higher readership, and greater attention from subsequent literature and, in some cases, from the general and social media \cite{intro_5,intro_6,intro_7,intro_8}. 
It is important to note that this phenomenon is not unique to academic publications and, in fact, it has been well documented for news pieces \cite{news}, literature \cite{lit}, and social media posts \cite{social_media_style}, to name a few.  
Research has shown that, in many cases, scholars present a unique writing style (WS), resulting in impressively accurate authorship classification and profiling algorithms \cite{author_change,author_change_2,author_change_3,author_change_4,journal_1}. 
However, as a collaboration among scholars has become increasingly prevalent in modern science \cite{amjad2017standing,wuchty2007increasing,journal_5}, so did the practice of collaborative writing \cite{credit,bozeman2004scientists,zhang2018understanding}, resulting in many co-authored manuscripts having a \say{mixed style}. Namely, the individual style of each scholar is reflected differently, and to a different extent, in the resulting co-authored manuscripts \cite{mixed_style_1,mixed_style_2}. 

Common to the literature in this realm is the focus on \textit{manuscripts} as the unit of analysis \cite{journal_3}. In other words, past research has predominantly considered each manuscript separately, analyzing its content and/or its' (co-)authors' identity or characteristics. However, research in WS considering  \textit{scholars} as the unit of analysis is significantly less prevalent. 

WS changes have been extensively investigated from a  social, cultural, and professional perspective \cite{ws_general_1,ws_general_2,ws_general_3,ws_general_4,ws_general_5}. For instance, \cite{ws_rw_1} developed a computational model that can detect authors of online messages. The authors show that parameters such as age, gender, and mother tongue are strong indicators, on average, for one's WS. In a similar manner, \cite{ws_rw_2} show that gender is statistically associated with the WS of individuals in multiple types of texts. Moreover, \cite{ws_rw_3} show that, at the same point in time, authors from different ages have different WS. \cite{ws_rw_4} extended this line of work, showing that the same individuals changed their WS as they grow up from children to adults. To the best of our knowledge, an investigation into how a scholar's academic WS is evolving and shaping throughout one's career, especially considering its academic relations with his/her advisors and peers, has yet to be examined in the literature. 

In this work, we seek to address the following key questions:

\begin{itemize}
    \item How does one's WS significantly change over time? 
    \item How do research students (i.e., advisees) part from their advisors' WS?
    \item How is a scholar's WS affected by her collaborations?  
\end{itemize}

To answer these questions, we develop a computational methodology combining a temporal graph representation of co-authorship dynamics, natural language processing, and deep learning techniques. We apply our methodology to real-world, large-scale bibliographic data from the Computer Science (CS) discipline consisting of around 570 thousand CS scholars and 13.7 million indexed academic publications. Our results show that authors' WS changes over time with some form of \say{WS convergence} occurring around one's 13th published manuscript. Similarly, scholars' WS seem to part from their academic advisors' WS early on in their careers, stabilizing around the 14th published manuscript. Finally, several factors such as gender and field of research are found to statistically associate with one's WS changes due to collaborations.  

The remainder of this manuscript is organized as follows: Section \ref{sec:methods}, details the methods and data used in this work. Section \ref{sec:results} outlines the results obtained from our analysis followed by their discussion in Section \ref{sec:discussion}. Finally, Section \ref{sec:conclusions} draws conclusions, and highlights possible future work avenues.

\section{Methods and Materials}
\label{sec:methods}

\begin{figure}
    \centering
    \includegraphics[width=0.99\textwidth]{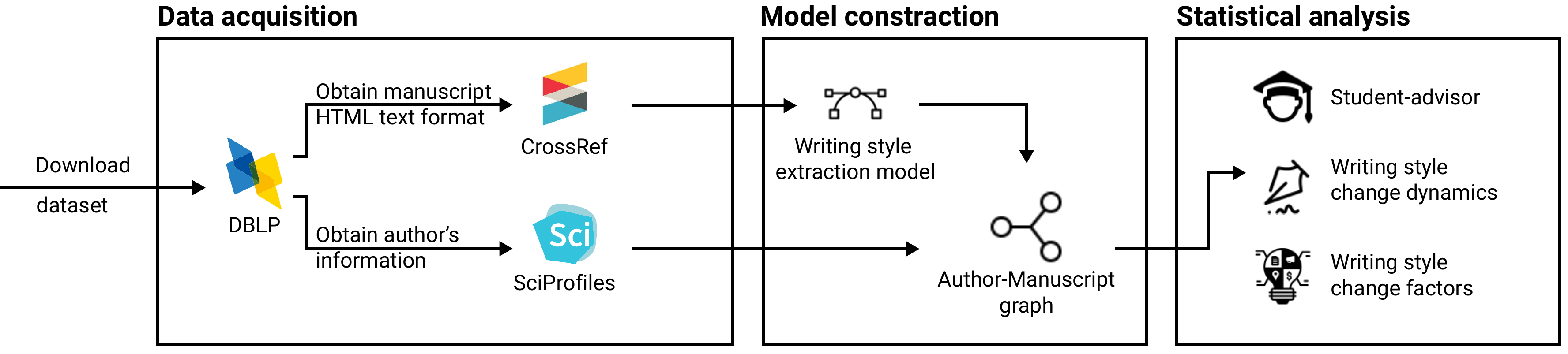}
    \caption{A schematic view of the study's methodology.}
    \label{fig:methods}
\end{figure}

Our methodology consists of three phases: First, we rely on the extensive CS literature indexed by the popular DBLP dataset \cite{dblp} and retrieve the original manuscripts and author profiles from CrossRef\footnote{https://api.crossref.org/} and SciProfiles\footnote{https://sciprofiles.com/}, respectively. 
Second, the acquired data is used to populate a time-depended social graph-based mathematical model.
Third, the resulting non-trivial model is analyzed to address the three research questions introduced above.  Fig. \ref{fig:methods} presents a schematic view of the study's methodology. 

\subsection{Data acquisition}
\label{sec:data_acquisition}

The DBLP database is a specialized bibliographic database that provides open bibliographic information on major CS journals and conference proceedings with decent coverage and accuracy \cite{rosenfeld2023dblp}.
The database was retrieved on March 20th, 2023 resulting in 14,301,639 publications. Using their DOIs (Digital Object Identifiers), 13,874,575 (97\%) publications were matched to their original text using CrossRef and fully recovered in an HTML format. These publications were authored by 610,281 different authors. Using SciProfiles, 593,513 (97.2\%) were matched and their profiles were retrieved.
We filter out scholars who published over 500 manuscripts (less than 1\%). 
Overall, more than 13.7 Million CS publications, authored by more than 587 Thousand scholars  are considered in the following.
On average, each author published \(23.42 \pm 40.44\) manuscripts in our data.

\subsection{Model construction}
\label{sec:text_style}
It is important to note that formally representing the WS of any given text is a challenging task as no clear, agreed-upon, definition of WS is currently available \cite{mixed_style_2}. As such, prior computational studies that dealt with large volumes of texts, which are self-evidently infeasible for domain experts to manually tag, adopted a data-driven approach \cite{nlp_good}. Specifically, deep-learning-based models are considered state-of-the-art in this realm \cite{dl_in_nlp}. For our study, we align with prior work and represent a manuscript's WS using a state-of-the-art model proposed by \cite{text_style_base}. The adopted model is based on the assumption that large pre-trained text-to-text artificial neural networks encompass the textual WS that can be used to condition the decoder of a style Transfer-based models \cite{transformer} through a fine-tuning procedure \cite{text_style_base}. Technically, a text-to-text transformer model based on the T5 architecture called \textit{TextSETTR} \cite{t5}, is adopted. TextSETTR generates a 1024-dimensional vector and accepts an arbitrarily long text using the attention layer in the transformer model. The original developers of the model have shown that the model outperforms the previous models in a wide range of settings. Of interest to our context, the model was shown to align with WS expert opinions more than 80\% of the time. This exceptional performance is obtained using the Few-shot machine learning approach \cite{few_shot} which, unlike other approaches such as supervised or unsupervised learning \cite{not_few_shot_1,not_few_shot_2}, requires very few labeled training examples during inference. Taken jointly, these properties make the proposed model seem especially suited for our research challenge.

By applying the \textit{TextSETTR} model to a given manuscript we are provided with a representation of that manuscript's WS. However, for our purposes, we are mainly interested in representing a \textit{scholar's WS}. Accordingly, if the manuscript is authored by a single scholar then that manuscript's WS can be fully attributed to the scholar alone at that time. However, if the manuscript is co-authored by several scholars, the resulting manuscript's WS is assumed to be a mixture of its co-authors' prior WSs. In order to disentangle this WS mixture, we assume that each part of the manuscript was written by a single scholar, as is commonly assumed in prior literature \cite{author_change_intro}. Accordingly, we perform the following process: First, we divide the manuscript into textual components such that each one presents a distinct, yet consistent, WS. To that end, we utilize the state-of-the-art model proposed by \cite{author_change} which demonstrated 85\% accuracy for this task on a large volume of documents. Once the division into textual components is obtained, we use the \textit{TextSETTR} model discussed above to get a vector representation of each component in the text separately. Finally, we map each of the resulting WS vectors to their assumed source (i.e., scholar) by matching each vector to the co-author who is currently represented by the most similar WS vector using a standard Euclidean distance metric\footnote{Unlikely ties may result in a single component being assigned to multiple co-authors.}. If a scholar is associated with a single component, then that component's WS is treated as that scholar's WS in that point in time. However, if more than a single component is associated with a scholar, then the proportional average of the components' WS vectors is used instead. Clearly, to perform the last step (i.e., components to scholars mapping), each co-author's prior WS is needed. Assuming a scholar has at least one single-authored manuscript, that manuscript's WS can be used as a starting point for our iterative procedure. Specifically, starting from a scholar's first solo-authored manuscript, the identified WS at that time point is iteratively propagated to the proceeding and precluding manuscripts according to the procedure outlined above. In other words, starting at each scholar's first single-authored manuscript, we use the obtained WS to assign the components (and their WS) of the immediately proceeding and precluding manuscripts which, in turn, are used for their proceeding and precluding manuscripts, and so on. Scholars without any single-authored publications were omitted from further consideration (less than 3\%). 

\subsection{Author-Manuscript Graph}
\label{sec:network}
To capture the collaboration and WS dynamics over time, we define a graph, \(G = (S, M, E)\), where each scholar is represented as a node in the graph \(s \in S\), each manuscript is represented as a different type of node in the graph \(m \in M\), and a directed edge \(e \in E \subseteq A \times P\) connects each scholar to each of his/her manuscripts. Formally, a scholar node \(s \in S\) is defined by the tuple \(s := (f, g) \) where \(f\) is the scholar's main field of the study as indicated by the name of its primary-associated department (e.g., computer science, mathematics, physics)\footnote{We extracted this information by searching the scholar's name in Google and retrieving the data from the first link, followed by a manual regular expression data standardization.} and \(g\) is the scholar's gender that can take one of the values \(\{male, female, unknown\}\). The scholar's gender is obtained according to a query to the model proposed by \cite{gender_model} which was trained on around 100 million pairs of names and gender association, as collected by \textit{Yahoo!}. A scholar's gender is taken only if the model's prediction confidence is higher than 95\% (true for 94.6\% of the scholars). A manuscript node \(m \in M\) is defined by the WS vectors associated with the manuscript's components and the time the manuscript has been published, denoted by \(\xi\) and \(t\), respectively. Last, an edge \(e = (s, m) \in E\) indicates that a scholar \(s\) is a (co-)author of the manuscript \(m\). Overall, we find the main field of study of \(81.3\%\) of the scholars and the gender of \(94.6\%\). In total, 80.5\% of scholars' profiles included both parameters.

\subsection{Statistical analysis}
\label{sec:statistical_analysis}
The statistical analysis is divided into three parts, each corresponding to one of our primary research questions. 

\paragraph{WS Dynamics:} To quantify the WS change over time for a given scholar, we define a function \(C: \mathbb{R}^{\kappa \times N} \times \mathbb{R}^{N} \rightarrow \mathbb{R}^+\) where \(\kappa\) is the yearly average number (rounded to the closest natural number) of papers an author publishes in a year. In addition, \(N \in \mathbb{N}\) is the WS representation vector's dimension. Formally, \(C\) accepts a list (\(L\)) of \(\kappa \in \mathbb{N}\) WS vectors corresponding to \(\kappa\) WS vectors published before a reference WS, \(u\), which is also provided to \(C\): 
\begin{equation}
    C(L, u) := || \frac{1}{\kappa} \sum_{v \in L} (v) - u||,
    \label{eq:ws_change_over_time}
\end{equation}
where \(||x||\) is the \(L_1\)  norm of a vector \(x\). Intuitively, the above calculation quantifies the extent to which a currently exhibited WS is different compared to former WSs presented roughly during the preceding year.  

\paragraph{WS Emergence:} To capture how research students' WS emerges, one needs to determine who were a student's advisors and when that student graduated. While some crowd-sourced advisor-advisee data is available by the Mathematics Genealogy Project\footnote{\url{https://www.genealogy.math.ndsu.nodak.edu/}} and Academic Family Tree\footnote{\url{https://academictree.org/}}, from our preliminary investigation, it does not cover a significant portion of our data. As such, we adopt a heuristic approach which was successfully applied in prior works (e.g., \cite{student_advisor_trick}) where we consider the individual(s) a scholar has co-authored the most manuscripts during their first three publications years as their advisor(s). Note that this simple heuristic may capture both \say{official} and \say{unoffical} advisors alike, which seems favorable for our purposes. Formally, let us denote the set of manuscripts the student and advisor(s) co-authored during the first three years to be \(A\) and \(\rho := |A|\in \mathbb{N}\). Thus, the student's exposure to the advisor's WS is set to be the average WS of the advisor from the manuscripts in set \(A\). Hence, the student's style emergence function with respect to his/her advisor(s) (\(A\)) is defined as follows: \(\delta_{A}(i): \mathbb{R}^{N} \rightarrow \mathbb{R}^{+} \) such that \(\delta_A(u) = || \frac{1}{\rho}\sum_{v \in A} (v) - u ||\), where \(u \in \mathbb{R}^N\) is the student's WS vector one wishes to compare with the baseline WS which is computed by \(\frac{1}{\rho}\sum_{v \in A} (v)\). Since we are interested in  the temporal change of a scholar's WS, let us define \(u_i\) to be a scholar's \(i_{th}\) WS vector such that \(u_0\) is the first manuscript the scholar published following the latest manuscript in \(A\). Intuitively, we compare a currently exhibited WS to the advisor's average WS presented to the student as part of their co-authored manuscripts during the student's training period. 

\paragraph{WS and Collaborations:} For each co-authored manuscript, we extract the following features:  for each co-author, we retrieve the main field of study (\(f\)), gender (\(g\)), and the number of previously published manuscripts. The number of co-authors listed in the manuscript's byline is also extracted. These values are considered as a feature vector \(x\) for our learning model. We define the scholar's WS change which was observed due to a co-authored manuscript as defined by Eq. (\ref{eq:ws_change_over_time}), to be the target value -- denoted as \(y\). Since the number of co-authors can be arbitrary, we set \(x\)'s size to be the maximal size required by any manuscript in the database and padded the non-required positions in \(x\) accordingly. 

The resulting dataset of samples, consisting of the features of each co-author as input and the observed WS change as an output, is fed to a Tree-based Pipeline Optimization Tool (TPOT) automatic machine learning model (AutoML) \cite{tpot,teddy_amit} that is especially suited for complex regression tasks and seeks to minimize  the predictions' mean absolute error. Feature importance is computed and reported to determine the perceived influence each parameter had on the prediction capability of the model \cite{feature_importance}. 

In addition, we classify each WS change (i.e., \(y\)) to one of the following types: 1) towards the center of mass (i.e., all co-authors' WS move closer to the average WS of the group); 2) positive one-side change (i.e., the scholar in question moves closer to the average WS of the group but the prior criterion is not met); 3) negative one-side change (i.e., the scholar in question moves away from the average WS of the group) and 4) no clear change (i.e., if no other criteria are met).
Accordingly, we perform a statistical analysis to determine if certain circumstances, as detailed in the following analysis, are statistically associated with different WS change types using an ANOVA test with post-hoc Tukey correction. 

\section{Results}
\label{sec:results}

\begin{figure}
    \centering
    \includegraphics[width=0.99\textwidth]{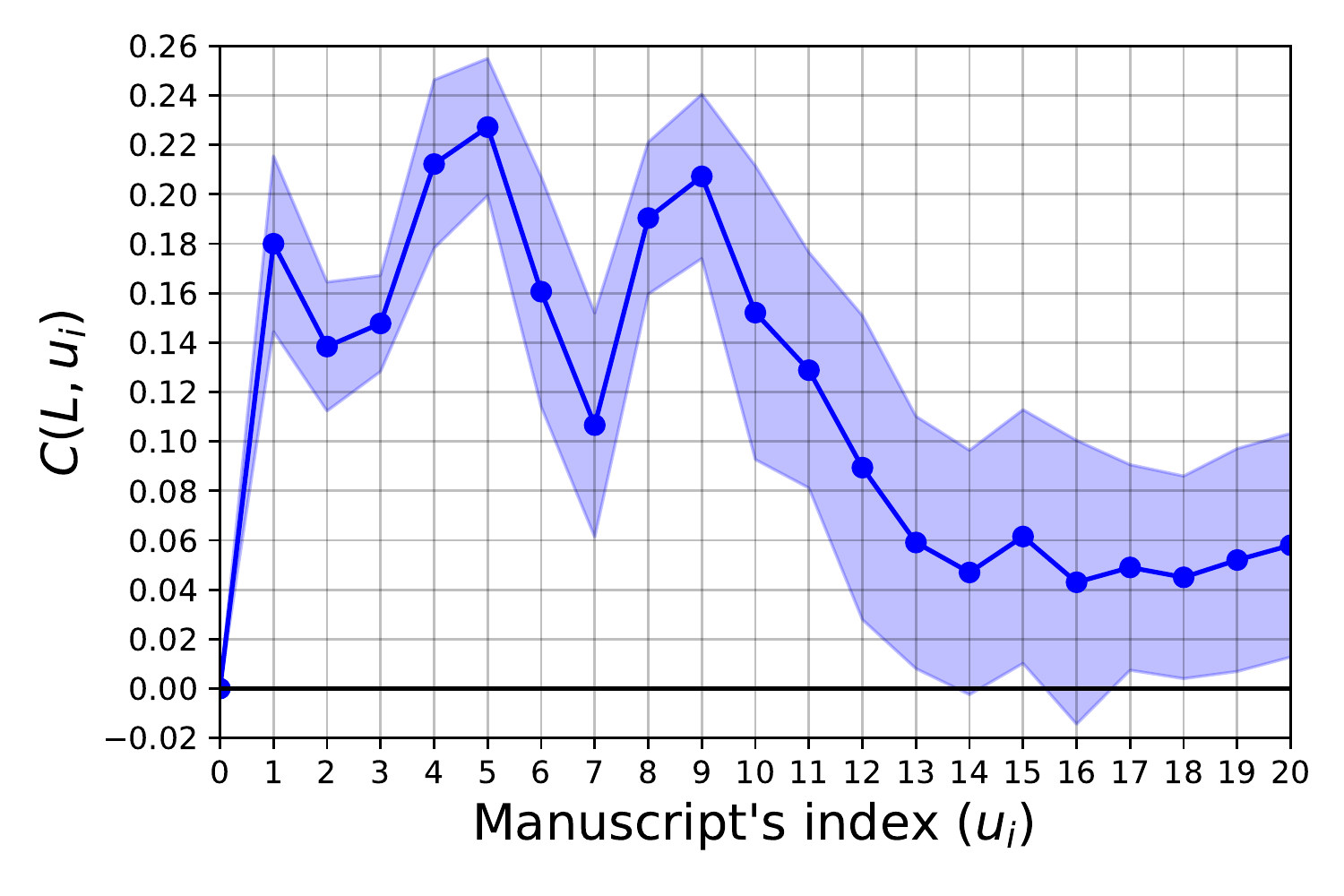}
    \caption{Mean \(\pm\) standard deviation of WS change for the entire studied population.}
    \label{fig:change_over_time}
\end{figure}

\paragraph{WS Dynamics:} Fig. \ref{fig:change_over_time} presents the authors' WS changes over manuscripts, where \(|L| = 3\). As one can notice, for roughly the first 12 published manuscripts, a scholar's WS varies significantly from one manuscript to the next. From that point on, the scholar's WS is \textit{not} constant (i.e., the WS change is not zero) yet the WS change seems to be mild and relatively stable over time. 
In order to mathematically capture this phenomenon, we introduce a threshold over the WS change to determine if and when a scholar's WS has converged. Formally, the convergence point, \(\alpha \in \mathbb{N}\), for a threshold \(\omega \in \mathbb{R}^+\), is defined as \(\alpha := \min_j \big ( \frac{1}{z-j}\sum_{i \in [j, \dots, z]} C(L, u_i) \big ) \leq \omega \). The results of this analysis are summarized in Table \ref{table:converage_analysis}. Note that since not all scholars converge for a given threshold  \(\omega\), we stated the percentage of scholars that did coverage given the specified threshold. 

\begin{table}[!ht]
\centering
\begin{tabular}{|l|c|c|c|c|c|}
\hline
\textbf{Threshold} (\(\omega\)) & \(0.01\) & \(0.02\) & \(0.03\) & \(0.04\) & \(0.05\) \\ \hline
\textbf{Convergance point} (\(\alpha\)) & \(13.3 \pm 6.7\) & \(10.8 \pm 4.9\) & \(8.5 \pm 4.2\) & \(7.1 \pm 3.6\) & \(5.2 \pm 3.0\) \\ \hline
\textbf{Convergence (\%)} & \(89.4\) & \(92.3\) & \(95.7\) & \(96.5\) & \(96.9\) \\ \hline
\end{tabular}
\caption{Mean \(\pm\) standard deviation of the convergence point (top row) and the percentage of converging scholars (bottom row) for each  examined threshold (columns).}
\label{table:converage_analysis}
\end{table}

\begin{figure}[!ht]
    \centering
    \includegraphics[width=0.7\textwidth]{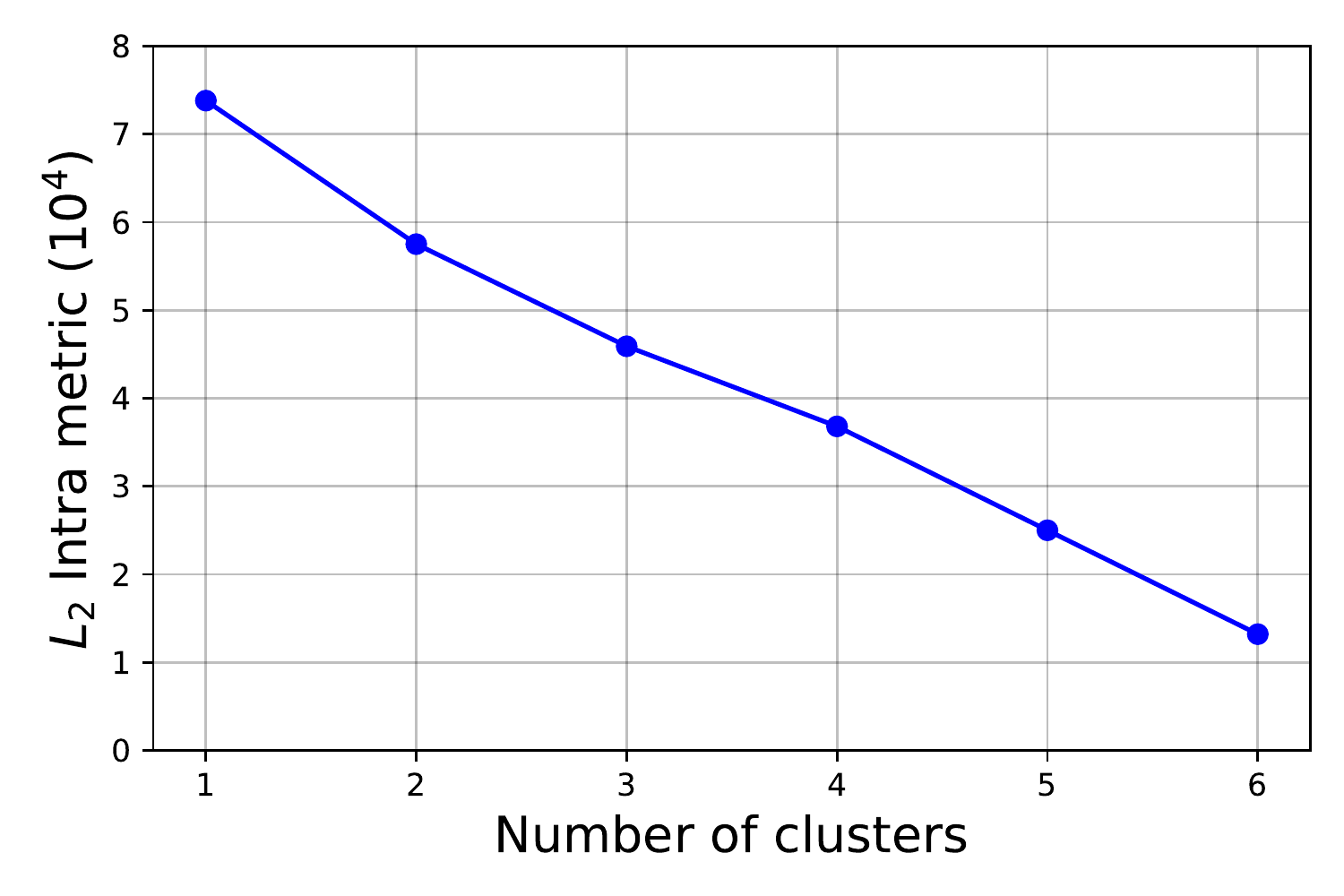}
    \caption{The Elbow graph of the k-means clustering method with the \(L_2\) distance. No apparent elbow point is observed.} 
    \label{fig:change_over_time_elbow}
\end{figure}

One may speculate that the WS change dynamics may be significantly different for different scholars. In order to examine this hypothesis, we used the classic k-mean algorithm adapted to time series data \cite{ts_k_means} using the popular \textit{tslearn} library \cite{tslearn}. Fig. \ref{fig:change_over_time_elbow} shows the \(L_2\) intra metric for a different number of clusters on our data. Commonly, when the data is inherently divided into \(k > 1\) clusters, one expects to witness an \say{elbow} in the graph which reveals a point in which the decrease in the intra metric changes from large to small. However, as shown in the figure,  this is not the case here. Hence, the data does not seem to support this hypothesis.

\begin{figure}
    \centering
    \includegraphics[width=0.99\textwidth]{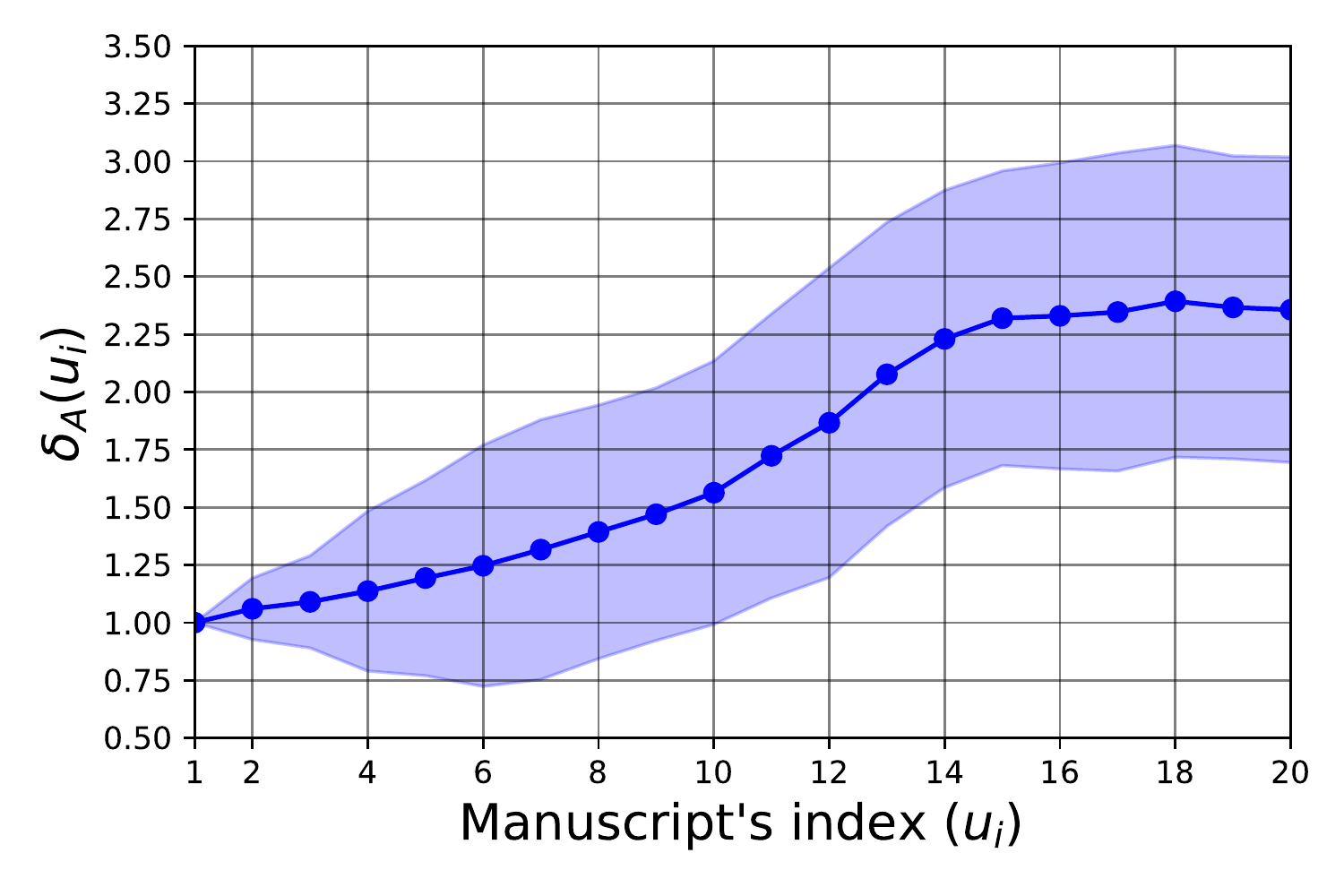}
    \caption{Mean \(\pm\) standard deviation of the difference between a scholar's WS and his/her advisors over manuscripts.}
    \label{fig:student_advisor_change_over_time}
\end{figure}

\paragraph{WS Emergence:}
Fig. \ref{fig:student_advisor_change_over_time} presents the students' WS emergence from the student's graduation. 
The figure depicts a sigmoid-like increase in the WS  difference from one's advisor(s) over manuscripts. Similarly to the analysis of WS change over time, roughly around the graduate's 14th publication, the difference from one's advisor(s) seems to converge to a relatively steady distance.

\begin{figure}[!ht]
    \centering
    \begin{subfigure}{.49\textwidth}
        \includegraphics[width=0.99\textwidth]{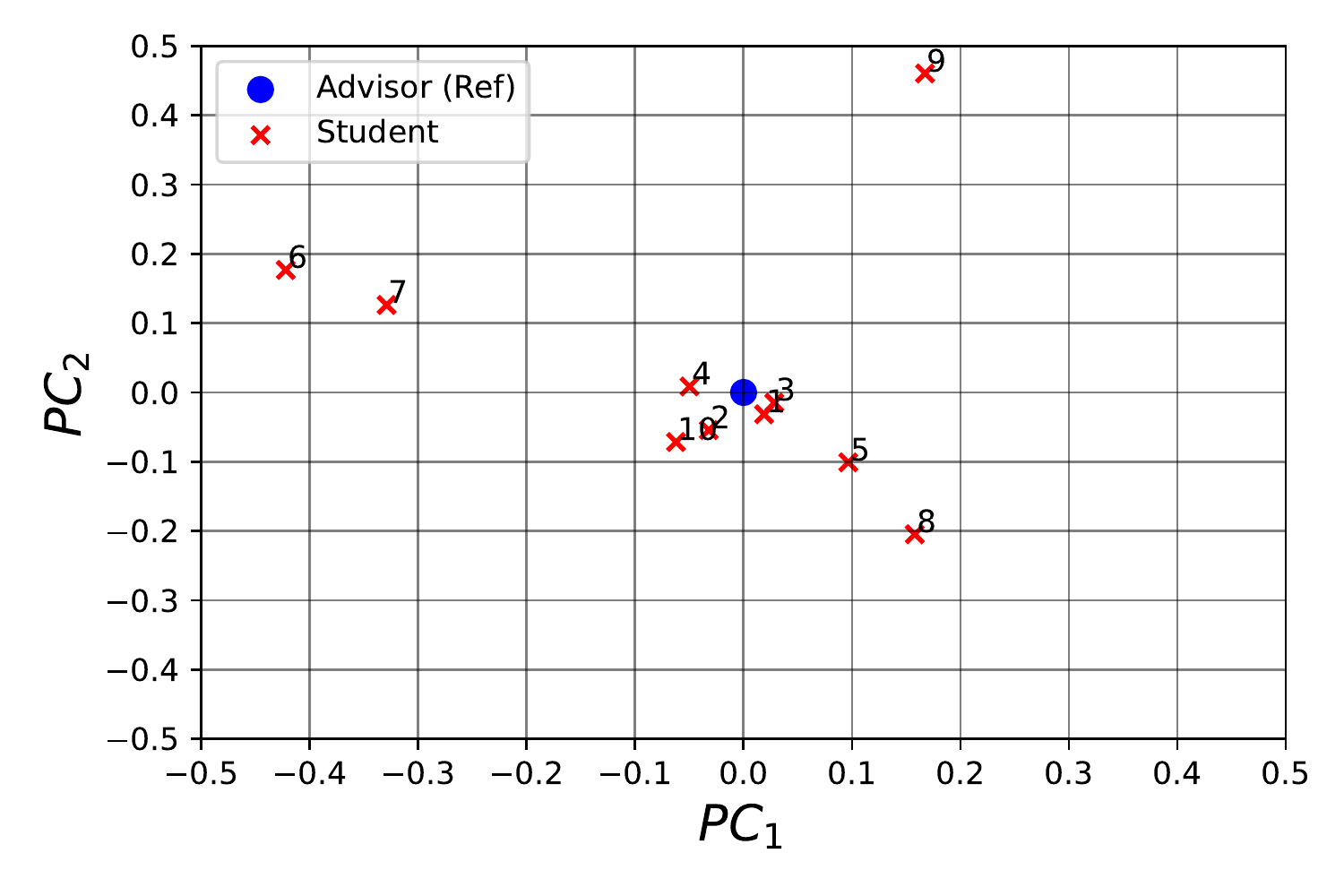}
        \caption{First author (\(\rho = 5\)).}
        \label{fig:student_advisor_teddy}
    \end{subfigure}    
    \begin{subfigure}{.49\textwidth}
        \includegraphics[width=0.99\textwidth]{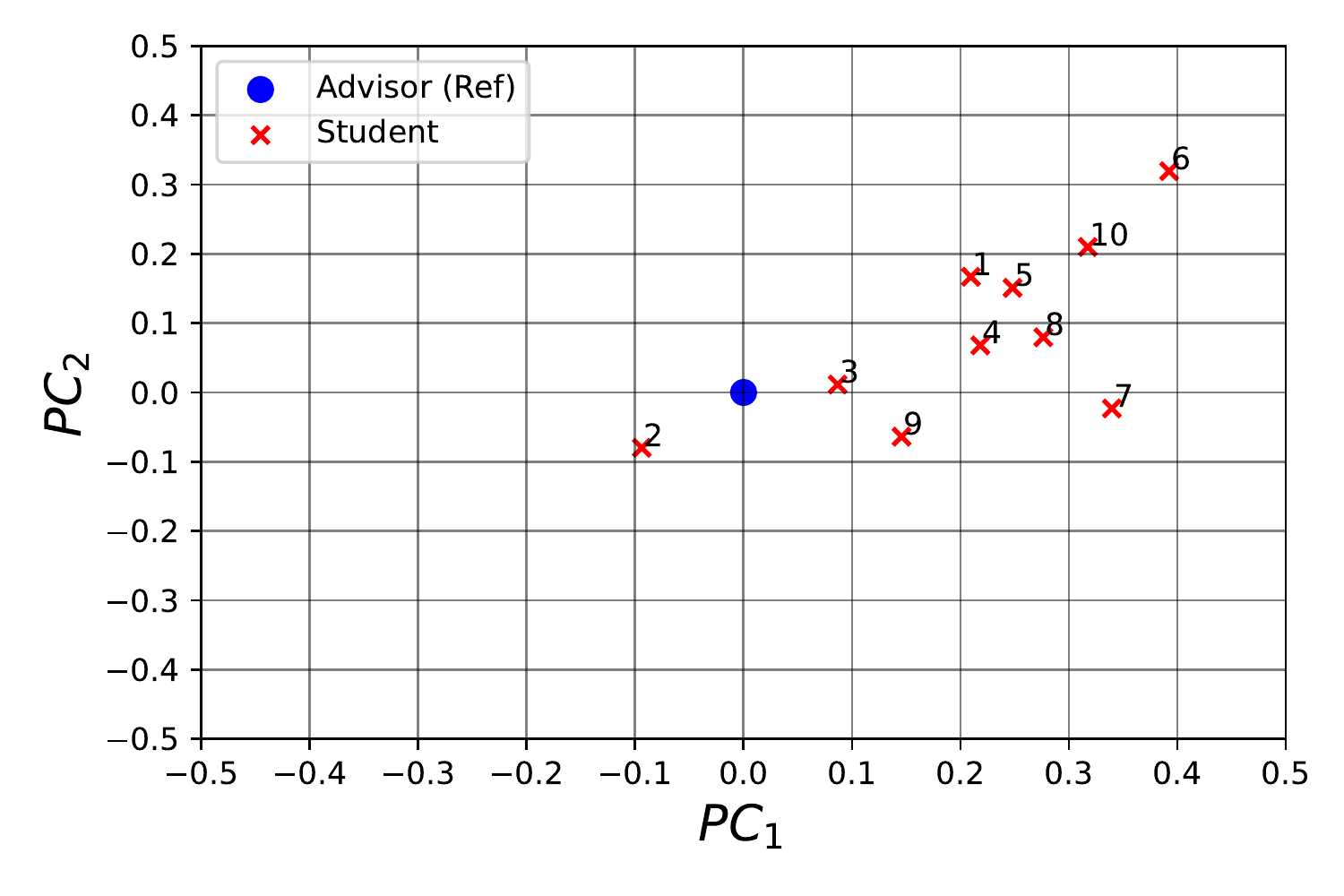}
        \caption{Second author (\(\rho = 12\)).}
        \label{fig:student_advisor_ariel}
    \end{subfigure}
    \caption{Scholars' manuscripts published after graduation are marked by 'x' and numbered by order of publication. }
    \label{fig:student_advisor_examples}
\end{figure}

Let us consider the two authors of this paper as illustrative examples. Fig. \ref{fig:student_advisor_examples} presents a 2-d standard PCA dimensionality reduction  projection of each of their first 10 publications after graduation compared to their respective advisors' WS during their training periods.  
The second author (presented on the right), demonstrates a rather consistent WS emergence pattern that moves away from his advisor's WS in the same direction over time. However, the first author (presented on the left) demonstrates a more cluttered pattern without a clear direction over time.   

\paragraph{WS and Collaboration:}
Since the number of features is considerably large due to the arbitrary number of co-authors in each  manuscript, we focused on the four factors summarized in Table \ref{table:change_factors}. Specifically, we first consider the genders involved in the co-authored manuscript, and for each examined scholar we classify each co-authored manuscript into one of four categories: 1) All male (i.e., all co-authors are male); 2) Male-Mix (i.e., the scholar in question is Male and the remaining co-authors consist of both male and female co-authors); 3) Female-Mix (i.e., the scholar in question is Female and the remaining co-authors consist of both male and female co-authors); 4) All female (i.e., all co-authors are female). We then consider whether the field of research of all co-authors is the same or not (acknowledging minor differences such as word order  and the insignificance of generic terms such as \say{department} or \say{faculty}). The \say{number of co-authors} factor refers to the number of authors listed in a manuscript's byline and the previous publications factor refers to the number of prior publications made by the scholar in question.     
For each of these factors, the average importance of changing one's WS is presented. In addition, the statistical relation between each value and a specific type of WS change is reported. 

\begin{table}[!ht]
\begin{tabular}{p{0.25\textwidth}p{0.2\textwidth}p{0.1\textwidth}p{0.35\textwidth}}
\hline
\textbf{Examined Factor} & \textbf{Average importance} & \textbf{Sample size} & \textbf{Typical change direction} \\ \hline
Gender & \(0.11\) &  \begin{tabular}[c]{@{}l@{}} 15M  \\ 32M \\ 21M \\ 4M \end{tabular} & \begin{tabular}[c]{@{}l@{}}\textit{All male}: towards center of mass*\\ \textit{Male-Mix}: no change** \\ \textit{Female-Mix}: no change* \\ \textit{All Female}: positive one side change* \end{tabular}
 \\\hline
 
Field of research & \(0.23\) &  \begin{tabular}[c]{@{}l@{}} 29M  \\ 43M \end{tabular} & \begin{tabular}[c]{@{}l@{}}\textit{Identical}: towards center of mass**\\ \textit{Different}: no clear change** \end{tabular} \\
\hline
Number of co-authors & \(0.32\) &  \begin{tabular}[c]{@{}l@{}} 9M \\ 17M \\ 46M \end{tabular}  &  \begin{tabular}[c]{@{}l@{}}\textit{2}: towards center of mass\\ \textit{3}: positive one side change* \\ \textit{4+}: no clear change** \end{tabular} \\
\hline
Previous publications & \(0.34\) &  \begin{tabular}[c]{@{}l@{}} 2M \\ 4M \\ 37M \end{tabular}  & \begin{tabular}[c]{@{}l@{}}\textit{1-3}: positive one side change*\\ \textit{4-13}: towards center of mass*\\ \textit{14+}: no clear change* 
\end{tabular}
\\
 \hline
 \end{tabular}
\caption{For each examined factor (rows) we report the estimated  importance in explaining the WS change observed as a result of a joint publication and the  statistical association with a specific type of WS change. Statistically significant results are  marked by * for \(p \leq 0.05\) and ** for \(p \leq 0.01\).}
\label{table:change_factors}
\end{table}

\section{Discussion}
\label{sec:discussion}
Let us revisit the original research questions posed for this study. 

First, we have asked \say{How does one’s WS significantly change over time?}. The results seem to indicate that the vast majority of scholars exhibit an evolving WS which, at first, presents a cluttered behavior that soon converges to mild and steady changes around their 13th manuscript. The fact that one's WS converges to small and steady changes between one manuscript to the next is somewhat intuitive as it reflects the process of forming one's unique academic personality, style, and practices which are ever-evolving. However, the fact that this convergence occurs early in one's career is, to us, very surprising. In our data, on average, convergence occurs after four publication years. This means that the scholars' WS \say{learning curve} has flattened extremely early. One possible explanation may be the infamous pressure to publish extensively during one's first years in academia, partially, to secure a permanent position \cite{publish_hard}. Specifically, during these first years a scholar may avoid the long, arguably needed, process of perfecting their WS in exchange for improving their body of work.   
%

Second, we have asked \say{How do research students (i.e., advisees) part from their advisors’ WS?}. The results seem to suggest that the distance between one's WS and his/her advisors' WS is increasing in a sigmoid-like fashion until convergence is reached around one's 14th publication. 
Interestingly, this convergence seems to agree with the one obtained from the previous analysis as well. The observed pattern seems to align with the historically observed dynamics of apprenticeship \cite{discussion_1}. The results also demonstrate an increasing pattern in standard deviation presented in Fig. \ref{fig:student_advisor_change_over_time}. These indicate that one's departure from his/her advisors' WS is very personal, aligning with the results of a recent study dedicated to the advisor-advisee collaboration patterns in Computer Science \cite{rosenfeld2022should}. 

Last, we have asked \say{How is a scholar’s WS affected by her collaborations?}. The results point to several statistically significant factors that seem to govern the way collaborations influence one's WS. Starting with gender, it is found to be the least influential factor out of the examined ones. Interestingly, while the interactions between males and females are symmetric in the sense that, statistically, they do not have a specific way of changing one's WS, interactions between the same gender result in different outcomes. This outcome agrees with a wide range of prior studies about collaborations and gender \cite{discusison_gender_1,discusison_gender_2,discusison_gender_3}. The field of research seems to play a slightly more central role. Specifically, co-authors from the same field are well influenced by each other's WS while co-authors from different fields are not. This result is, perhaps, counter-intuitive as one could expect scholars from different disciplines to have a greater impact on each other as they are accustomed to slightly different writing standards and practices. However, this result is similar in spirit to how scholars react and adopt ideas from peers within and outside their research field  \cite{network_rumers}. In addition, the number of co-authors encompasses great importance in predicting the WS change due to collaboration. Albeit statistically insignificant, when there are only two co-authors, both seem to learn from each other and slightly adopt each other's WS. However, when three co-authors are concerned, and especially when more than three are considered, scholars are less influenced by their co-authors' WS.  The most significant factor is the previous number of manuscripts a scholar has. Similar to the results discussed earlier for the student-advisor dynamics, young scholars (in terms of published manuscripts) tend to be more influenced by others' WS compared to more experienced authors which have already established a personal WS and thus are less prone to changes.

\section{Conclusions}\label{sec:conclusions}

In this study, we explored how scholars' writing styles evolve throughout their careers focusing on their academic relations with their advisors and peers.
To this end, we proposed and implemented a computational framework that captures how scholars' WS changes over time due to co-authorship and advisor-advisee relationships. 
The obtained results point to several intriguing phenomena which, in addition to their fundamental role in understanding academic WS dynamics, can be instrumental in designing better Ph.D. and young faculty programs. 

It is important to note that the proposed model and analysis are not without limitations. First, our analysis focuses on the Computer Science discipline. In future work, we intend to extend our analysis to include additional disciplines which need not necessarily align with the practices and standards of Computer Science (e.g., Humanities and Social Sciences). Second, several parts of our implementation, such as the co-author's components mapping, are inaccurate almost by definition. Thus, the raw results used for our analysis are not without noise and errors. Improving these components could lead to more robust outcomes and conclusions. Finally, the proposed analysis does not take into consideration additional social and cultural features that might also govern scholars' WS and its changes \cite{liza}. For example, a scholar's nationality may likely play a central role in shaping his/her WS and its dynamics. We intend to explore this and additional socio-demographic features in the future.

\section*{Declarations}
\subsection*{Funding}
This research did not receive any specific grant from funding agencies in the public, commercial, or not-for-profit sectors.

\subsection*{Conflicts of interest/Competing interests}
None.

\subsection*{Data availability}
The data that has been used is presented in the manuscript with the relevant sources.
 
\bibliography{biblio}
\bibliographystyle{unsrt}

\end{document}